\documentclass[10pt, english]{article}
\usepackage[a4paper, margin=2.7cm]{geometry}
\usepackage[T1]{fontenc}
\usepackage[utf8]{inputenc}
\usepackage{babel}
\usepackage{authblk}
\usepackage{blindtext}

\usepackage{babel}

\usepackage{amsmath}		
	\usepackage{amssymb}
\usepackage{bbold}
           \usepackage{epsfig} 
	\usepackage{dsfont}      
	\usepackage{fixmath}     
	\usepackage{ifthen}
	\usepackage{epstopdf}
   \usepackage{appendix}

          \usepackage{enumerate}

           \usepackage{enumitem}

\usepackage{abstract}

	\usepackage{braket}   

  \newcommand*{\defeq}{\mathrel{\vcenter{\baselineskip0.5ex \lineskiplimit0pt
                     \hbox{\scriptsize.}\hbox{\scriptsize.}}}%
                     =}

  \newlength{\bracewidth}

\newcommand*\density[1]{\hat{\rho}_{#1}}

\newcommand{\aref}[1]{\hyperref[1]{Appendix~\ref{#1}}}
\newcommand{\difd}{\mathrm{d}}

\newcommand{\avg}[1]{\left\langle #1 \right\rangle}

\newcommand{\abssq}[1]{\left| #1 \right|^2}

\newcommand{\Ccurly}{\mathcal{C}}
\newcommand{\Tcurly}{\mathcal{T}}

\newcommand{\Gmc}{\mathcal{G}}

\title{{\bf \huge{Spatial interference between pairs of disjoint optical paths with a single chaotic source}}}

\author[1]{Michele Cassano}
\author[1,2]{Milena D'Angelo}
\author[1,2]{Augusto Garuccio}
\author[3,4]{Tao Peng}
\author[3]{Yanhua Shih,}
\author[5,6]{Vincenzo Tamma \thanks{vincenzo.tamma@port.ac.uk}}
\affil[1]{Dipartimento Interateneo di Fisica, Universit\`{a} degli Studi di Bari, 70100 Bari, Italy}
\affil[2]{Instituto Nazionale Di Fisica Nucleare, sez. di Bari, 70100 Bari, Italy}
\affil[3]{Department of Physics, University of Maryland, Baltimore County, Baltimore, MD 21250, USA}
\affil[4]{Institute for Quantum Science and Engineering, Texas A \& M University, College Station, Texas 77843, USA}
\affil[5]{Institut f\"{u}r Quantenphysik and Center for Integrated Quantum Science and Technology (IQ\textsuperscript{ST}), Universit\"{a}t Ulm, D-89069 Ulm, Germany}
\affil[6]{Faculty of Science, SEES, University of Portsmouth, Portsmouth PO1 3QL, UK}

\begin{document}
\maketitle

\begin{abstract}
We demonstrate a novel second-order spatial interference effect between two indistinguishable pairs of disjoint optical paths from a {\it single} chaotic source. Beside providing a deeper understanding of the physics of multi-photon  interference and coherence, the effect enables retrieving information on both the spatial structure and the relative position of two distant double-pinhole masks, in the absence of first order coherence. We also demonstrate the exploitation of the phenomenon for simulating quantum logic gates, including a controlled-NOT gate operation.
\end{abstract}



\section{Introduction}

The second-order interference phenomena investigated in the mid-1950s by Hanbury Brown and Twiss (HBT) imposed a deep change in the understanding of interference and coherence \cite{HBT1, HBT2}. In fact, the intense debate raised by HBT interferometry naturally led to the development of quantum optics \cite{HBT1, HBT2, Glauber1963, GlauberLecture}, with its intriguing fundamental studies on multi-photon interference \cite{ShihAlley, HOM, HOMspatial, LiuSpatial, tammalaibacher2014a, tammalaibacher2014,Tam-Sei}, its promising applications (e.g., imaging  \cite{D'AnCheShi, Lemos, Jane, PittShi, ValScaD'AnShi, D'AnShi, Scarcelli,  gatti2, ferri, Chen, Luo, Genov}, quantum information processing \cite{Kok, nielsen,tamma2014, laibachertamma2015, tammalaibacher2015a, tammalaibacher2015b}, metrology \cite{GioLloMac1, GioLloMac, D'AnCheShi, Dowling}, etc.), and developments (e.g., N-photon state characterization \cite{tammalaibacher2014, Legero}, entanglement generation  \cite{tammalaibacher2014, NOON} and entanglement simulation \cite{PhysRevA.57.R1477, PhysRevA.63.062302, Lee2002, Kagalwala2013, PengShih}).

In the original HBT interferometer \cite{HBT1}, second order interference is observed when light emitted by a single chaotic source  is detected by two separate sensors and correlation measurements are performed while varying either the time delay between the two detectors (temporal second-order interference) or their relative position (spatial second-order interference). The two detectors, separately, do not retrieve any (first-order) interference. However, interference is observed at second-order provided the time delay and the spatial separation are within the coherence time and the coherence area of the source, respectively.

Recently, Tamma and Seiler have proposed a modification of this scheme \cite{Tam-Sei}: before reaching the detectors, chaotic light propagates though two unbalanced M-Z interferometers. No first-order interference exists at the exit of the interferometers, since the unbalancing is larger than the coherence length of the source. Interestingly, interference between two long and two short paths is predicted to occur even if the relative time-delay between the two pairs is beyond the coherence time of the source. This interferometer, substantially different from previous schemes based on multiple incoherent sources \cite{Agaf, Pearce2015, SVS, Oppel2012}, thus offers a deeper insight on the interplay between interference and coherence in multiphoton interferometry. Furthermore, a controlled-NOT (CNOT) gate operation \cite{pittman, O'Brien, sanaka} can be simulated by employing this interference effect \cite{Tam-Sei}.

In this paper, we demonstrate that pure second-order interference between pairs of disjoint optical paths {(paths which do not overlap spatially)}, originated from a single chaotic source, can also be  observed in the spatial domain. The uniqueness of such a  spatial interference phenomenon stands in its potential application for sensing of remote objects. In particular, we consider an optical interferometer (Fig. \ref{C-NotPoP1a}) where the light from a single chaotic source, after being split by a balanced beam splitter, propagates through two double-pinhole masks placed in the two separate output channels of the beam splitter. The separation between the pinholes in each mask is such that no first-order interference can be observed by the detectors placed behind each mask. However, as shown in Section \ref{BasicSetup},  by measuring the correlation between the photon number fluctuations  at given transverse positions of the two detectors, a spatial second-order interference is predicted to appear. Interference occurs between  two pairs of disjoint optical paths, which are defined by the  two pairs of pinholes ($1_C,1_T$) and ($2_C,2_T$). In Section \ref{Applications}, we show that the information about the spatial structure and the relative position of the two masks is encoded within the relative phase between the two pairs of interfering paths, independently of the distance between the two masks and the source. In particular, we demonstrate that: 1) this information can be retrieved in suitable experimental  scenarios {(Tables \ref{SensApp} and \ref{Tab2})};  
2) {the measurement precision can be increased by changing some experimental parameters rather than increasing the frequency of the light (Table \ref{SensApp} and Fig. \ref{ResMag})}. Finally, in Section \ref{SimCNOT}, we show that the proposed interference phenomenon can also be used to simulate quantum logic operations, including a CNOT gate.

{The novel spatial interference effect introduced in this paper has already triggered two experiments: 1) the experimental characterization of two remote double-slit masks within the experimental scenarios (v) in Table \ref{SensApp}, and (i), (ii) in Table \ref{Tab2} \cite{D'AnMaz}; 2) the experimental simulation of the CNOT-gate operation  based on the spatial interferometer introduced in Fig. \ref{C-NotPoP1b} \cite{PCTS}. The more general results reported here provide the complete physical picture of the novel interference effect, and are likely to inspire further theoretical and experimental works (e.g., monitoring the relative change in the spatial structure of two distant masks, as predicted in Fig. \ref{ResMag}). Intriguing applications in imaging and sensing of remote objects are in fact at reach with the current technology.}

\section{Spatial interference effect}\label{BasicSetup}

\begin{figure} [h!] 
\centering\includegraphics[width=14cm]{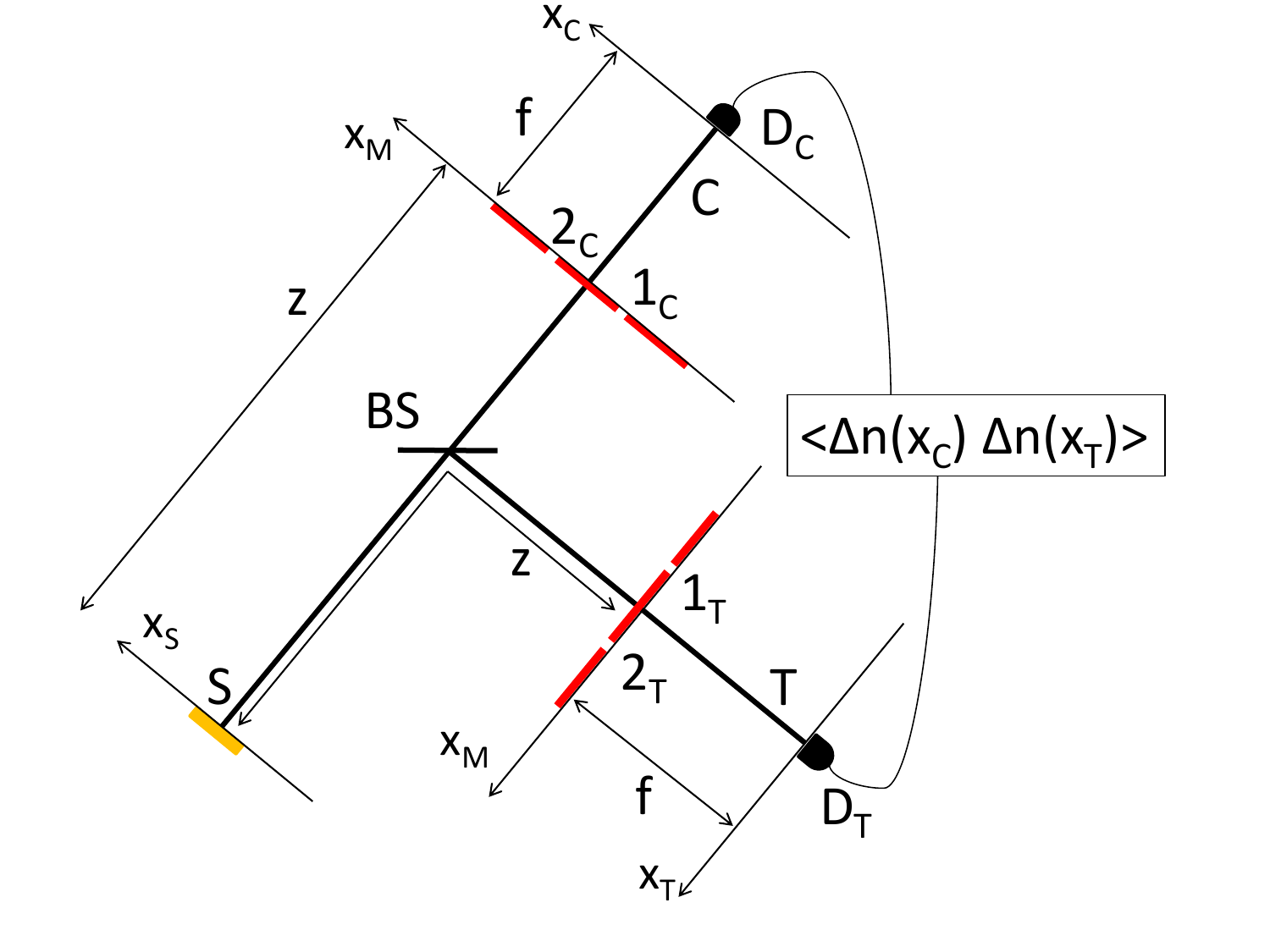}
\caption[]{Optical interferometer for sensing two remote double-pinhole masks through the observation of spatial second-order interference between indistinguishable pairs of disjoint optical paths.  Light emitted by a single 
 {1 dimensional} chaotic source, after being split by a balanced non-polarizing beam splitter, propagates through two double-pinhole masks placed at the same distance $z$ from the source and reaches two point-like detectors, $D_C$ and $D_T$, placed at distance $f$ from the masks. A correlation measurement between the fluctuations of the number of photons at the detectors $D_C$ and $D_T$ is performed.} \label{C-NotPoP1a}
\end{figure}

Let us start by introducing the interferometric setup depicted in Fig. \ref{C-NotPoP1a}: chaotic light emitted by the source S is split by a balanced non-polarizing beam splitter, and two double-pinhole masks are placed in the output ports of the beam splitter, at the same distance $z$ from the source. The pinholes are indicated as $1_C, 2_C$ for the upper mask and $1_T, 2_T$ for the lower mask. The light transmitted by the masks  reaches two point-like detectors, $D_C$ and $D_T$, placed at the same distance $f$ from the masks. A correlation measurement is performed between the fluctuations of the number of photons detected by $D_C $ and $D_T$. 

We first consider the correlation in the number of photons on the masks planes, which is given by  the second-order correlation function \cite{Glauber2007, Chen}
\begin{eqnarray}\label{G2defTh1}
G^{(2)}(x_p, x_{q})\propto\braket {n(x_p) n(x_{q})}=\braket {n(x_p)} \braket{n(x_{q})}+\braket {\Delta n(x_p) \Delta n(x_{q})},
\end{eqnarray}
with $p=1_C,2_C$ and $q= 1_T, 2_T$, where $n$ represents the photon number and $\Delta n \defeq n - \bar{n}$ the photon-number fluctuation around the mean $\bar{n}$. In particular, we consider the case of a quasi-monochromatic chaotic source, which, for simplicity, is also assumed to be 
 {1 dimensional} and linearly 
 {polarized} (e.g., along the horizontal H direction). 
The input chaotic light is described by \cite{Glauber2007, Mandel1995}
\begin{equation}\label{eq:Density_op_Thermal_B_Two_Mode0}
\density{H}= \int \left[ \prod_{\kappa} \difd^2 \alpha_{\kappa, H} \right] P(\{\alpha_{\kappa, H}\}) \bigotimes_{\kappa} \ket{\alpha_{\kappa, H}}_{S} \! \bra{\alpha_{\kappa, H}},
\end{equation}
with the Glauber-Sudarshan probability distribution \cite{Glauber1963, Sudarshan1963}
\begin{equation}
 P(\{\alpha_{\kappa, H}\})=\prod_{\kappa}\frac{1}{\pi\avg{n_{\kappa}}}\exp\left(-\frac{\abssq{\alpha_{\kappa, H}}}{\avg{n_{\kappa}}}\right),
\end{equation}
where $\alpha_{\kappa, H}$ are H-polarized coherent states, in the mode $\kappa$ associated with the $x$ component of the transverse wave vector,  and $\braket{n_{\kappa}}$ is the corresponding average photon number, which is assumed for simplicity to be constant \cite{Mandel1995}.
In this case, Eq. (\ref{G2defTh1}) reduces to \cite{D'AnShi}
\begin{eqnarray}
G^{(2)}(x_p, x_{q})&=&G^{(1)}(x_p) G^{(1)}(x_{q})+|G^{(1)}(x_p, x_{q})|^2,
\end{eqnarray}
where $G^{(1)}$ is the  first-order correlation function (see Eq. \ref{G1n}). Therefore, the second-order correlation function $G^{(2)}(x_p, x_{q})$ depends on two contributions: the first one, $G^{(1)}(x_p) G^{(1)}(x_{q}) \propto \avg{n(x_p)}\avg{n(x_q)}$, is a constant background; the second one, $\big{|} G^{(1)}(x_p, x_q) \big{|}^2 \propto \braket{\Delta n(x_p) \Delta n(x_{q})}$, is the interesting part of the correlation. The background can be removed  by performing  a correlation measurement between the fluctuations of the number of photons \cite{Chen}. The outcome of this measurement is different from zero for all the possible pairs of paths $(p,q) = (1_C, 1_T),  (2_C, 2_T), (1_C, 2_T), (2_C, 1_T)$, provided the relative distance between each pair of pinholes is smaller than the transverse coherence length  of the source ($l_{coh}$) on the plane of the masks, which is: $|x_p-x_q|\ll l_{coh}$. An interesting result comes out by working in the hypothesis that
\begin{enumerate}
\item the corresponding pairs of pinholes of the two masks are within the transverse coherence length, which is
\begin{eqnarray}
|x_{1_C} - x_{1_T}| \ll  l_{coh}    \label{Limit2b} \quad\quad
|x_{2_C} - x_{2_T}| \ll  l_{coh};
\end{eqnarray}
\item the pinholes, in each mask, are separated by a distance larger than the transverse coherence length of the source, which, given the condition {in Eq.} (\ref{Limit2b}), implies 
\begin{eqnarray}
|x_{1_C} - x_{2_T}| \gg  l_{coh} \label{Limit2a}  \quad\quad
|x_{1_T} - x_{2_C}| \gg  l_{coh}. 
\end{eqnarray}
\end{enumerate}
In fact, in this case, only the two pairs of paths ($1_C, 1_T$) and ($2_C, 2_T$), each one associated with two disjoint paths spatially coherent with respect to each other, contribute to the correlation, while no contribution comes from the two pairs of paths ($1_C, 2_T$) and ($2_C, 1_T$),
namely
\begin{eqnarray}
\braket {\Delta n(x_p) \Delta n(x_{q})}&\neq& 0
\Leftrightarrow (p,q) = (1_C, 1_T), (2_C, 2_T).
\end{eqnarray}
{Multi-photon correlations (``photon bunching'') thus give rise to the non-vanishing expectation value of the product of the photon-number fluctuations at the two remote pinholes $1_C$ and $1_T$  (or $2_C$ and $2_T$). This result arises from the correlation measurement and cannot be explained in terms of independent measurements at the two detectors. Interestingly, since the detectors are placed in the mask planes, the two pairs of disjoint paths ($1_C$,$1_T$) and ($2_C,2_T$) contribute independently of one another to the correlation measurement.}

What happens if we perform correlation measurements after the two-pinhole masks? Since light passing through the two pinholes of each mask is incoherent (condition {in Eq.} \ref{Limit2a}), one may expect that the two contributions  ($1_C, 1_T$) and ($2_C, 2_T$) add incoherently. However, as we shall show, they give rise to a counterintuitive spatial interference effect.
To demonstrate this result  we evaluate the correlation between the photon-number fluctuations $\Delta n(x_C)$ and $\Delta n(x_T)$ measured at equal detection times by the detectors $D_C$ and $D_T$, respectively, placed at the transverse position $x_C$ and $x_T$ behind the two-pinhole masks, namely
\begin{eqnarray} \label{Deltan} 
 \avg{\Delta n(x_C) \Delta n(x_T)}_{}&\propto& \big{|}G^{(1)}(x_C, x_T)\big{|}^2.
\end{eqnarray}
Here,
\begin{eqnarray} \label{G1n} 
G^{(1)}(x_C, x_T)&=& Tr [\density{H} \ \hat{E}_C^{{(-)}}(x_C) \hat{E}_T^{{(+)}}( x_T) ] 
\end{eqnarray}
is the first-order correlation function calculated at $x_C, x_T$, where $\hat{E}_d^+(x_d)$ and  $\hat{E}_d^-(x_d)$  are, respectively,  the positive and negative frequency part of  the electric field operator at the position $x_d$, namely
\begin{eqnarray}\label{Elecn} 
\hat{E}_{d}^{(+)}(x_{d})&=& K \int  d \kappa  g\{ \kappa; S, x_d\}  \hat{a}_S(\kappa),
\end{eqnarray}
where $K$ is a constant and $g\{ \kappa; S, x_d\}$ is the Green's function that describes the propagation of the mode $\kappa$ from  the source $S$ to the detector $D_d$, placed in $x_d$, with $d=C,T$,  and  $\hat{a}_S(\kappa)$ is the annihilation operator at the source S associated with the mode $\kappa$.

As demonstrated in Appendix \ref{app1}, in the paraxial approximation and by using the conditions given in {Eqs.} (\ref{Limit2b}) and (\ref{Limit2a}),  Eq. (\ref{Deltan}) becomes
\begin{eqnarray} \label{Deltan3} 
 &&\avg{\Delta n(x_C) \Delta n(x_T)}_{}\propto\big{|}G^{(1)}_{1_C,1_T}(x_C, x_T) + G^{(1)}_{2_C,2_T}(x_C, x_T)\big{|}^2, 
\end{eqnarray}
where $G^{(1)}_{1_C,1_T}$ and $G^{(1)}_{2_C,2_T}$ indicate the contributions to the correlation measurement coming from the two pairs of disjoint paths $(1_C, 1_T), (2_C, 2_T)$, respectively, and, as shown in Appendix \ref{appGF},
\begin{eqnarray} \label{G1eff10a} 
G^{(1)}_{p, q} (x_C, x_T) \propto B_p^*(x_C)  B_q(x_T) FT \left\{ |A(x_S)|^2 \right\} \left[ (x_p-x_q) /(\lambda z) \right],
\end{eqnarray}
with the two phase factors  $B_{p}^*(x_C)$ and $B_{q}(x_T)$ defined in Eq. (\ref{GreFunn21}) and the Fourier transform $FT \left\{ | A(x_S)|^2 \right\} \left[\chi \right]$  of the source  intensity profile $|A(x_S)|^2$ calculated at $\chi = (x_p - x_q) / (\lambda z)$. The result of Eq. (\ref{G1eff10a}) is at the core of the counterintuitive interference phenomenon addressed in this paper. In fact, it indicates that 
the contributions  $G^{(1)}_{p,q}$, associated with the pairs of paths (p,q) = ($1_C,1_T$), ($2_C, 2_T$), ($1_C, 2_T$), ($2_C,1_T$), strongly depend on  the relative distance $x_p-x_q$ between the remote pinholes $p$ (of mask $C$) and $q$  (of mask $T$) as compared to the transverse coherence length of the source  ($l_{coh}$) on the plane of the masks. In our scenario, due to the conditions {given in Eqs.} (\ref{Limit2b}) and (\ref{Limit2a}), we obtain
\begin{eqnarray}
G^{(1)}_{1_C,1_T}(x_C, x_T)\propto  B_{1_C}^*(x_C) B_{1_T}(x_T)&,& \ \ G^{(1)}_{2_C,2_T}(x_C, x_T)\propto  B_{2_C}^*(x_C) B_{2_T}(x_T)\nonumber
\\G^{(1)}_{1_C,2_T}(x_C, x_T)&=&G^{(1)}_{2_C,1_T}(x_C, x_T)=0. 
\end{eqnarray} 
Therefore, as reported in Eq. (\ref{Deltan3}), the correlation between the fluctuations of the number of photons  enables to retrieve the \textit{interference} between the 
 {two possible ``photon bunching'' contributions $G_{1_C,1_T}$  and $G_{2_C,2_T}$} associated with the pairs of disjoint paths $(1_C, 1_T)$ and $(2_C, 2_T)$. In fact, these two  {contributions add coherently and cannot be distinguished in the correlation measurement. As mentioned above, these two contributions can be distinguished when performing correlation measurements on the mask planes. In this case, these two  contributions lead to independent bunching events due to both the statistical properties of the chaotic source and the experimental conditions in Eqs. (\ref{Limit2b}) and (\ref{Limit2a}). In contrast, when correlation measurements are performed after the two masks, the two} 
 {pairs of path $(1_C, 1_T)$ and $(2_C, 2_T)$ become \textit{indistinguishable}. Multi-photon correlations emerge from the resulting interference between the two pairs of disjoint paths,} even if the pinholes in each mask are separated much further than the coherence length of the source.

\section{Sensing applications}\label{Applications}

As shown in Appendix \ref{app1}, the correlation in the fluctuation of the number of photons in Eq. (\ref{Deltan3}) can be written as
\begin{eqnarray}\label{Deltan4} 
 \avg{\Delta n(x_C) \Delta n(x_T)}_{} \propto \big{|}1+e^{  i \phi(s_{C}, d_C, s_{T}, d_T, x_C, x_T )}\big{|}^2, 
\end{eqnarray}
with
\begin{eqnarray}\label{phiP} 
\phi(s_{C}, d_C, s_{T}, d_T, x_C, x_T )= \frac{2 \pi }{\lambda} \Big(\frac{ s_{T} d_T - s_{C} d_C }{h}  
- \frac{x_{T} d_T - x_{C} d_C}{f}\Big),
\end{eqnarray}
where $h$ is defined by the condition $1/h=1/z+1/f$, $d_j \defeq x_{2_j}-x_{1_j}$ is the pinhole separation for the j-th mask and $s_j \defeq (x_{1_j}+x_{2_j})/2$ is the transverse coordinate of the center of the j-th mask, with $j=C,T$. Remarkably, the interference effect described by Eq. (\ref{Deltan4}) holds for any value of the parameters $z$ and $f$, namely, for any distance of the masks from the beam splitter and from the corresponding detectors.

\begin{table}
\begin{center}
\caption{
 {Summary of the conditions}  for  monitoring the transverse spatial structure and  position of two remote double-pinhole masks by 
 {performing the correlation measurement of Eq. (\ref{Deltan4}) in} the setup in Fig. \ref{C-NotPoP1a}. 
 {In each of the five experimental scenarios one variable parameter is monitored, and the other parameters are fixed in order to ``magnify''  the effect of small variations of the monitored parameter; the  corresponding ``magnification'' factors are reported in the third column of the table.} 
}
\begin{tabular}{|l|l|l|}
\hline
 Experimental conditions   & Variable  parameter   & 
 {``Magnification''}   \\  in addition to 
 {Eqs.} (\ref{Limit2b}) and (\ref{Limit2a}) &to monitor & 
  {factors} \\
\hline
{(i)} $x_T=x_C$, $s_{T}=s_{C}$  & $d_T-d_C$ & $s_C/h - x_C/f$           \\
{(ii)} $x_T=-x_C$, $s_{T}=-s_{C}$   & $d_T+d_C$  & $-s_C/h + x_C/f$       \\
{(iii)} $|s_T| \neq |s_C|$   & $d_{T,C}$  & $s_{T,C}/h - x_{T,C}/f$       \\
{(iv)} $d_T=d_C$  & $s_{T}-s_{C}$  & $d_C/h$          \\
{(v)} $d_T \neq d_C$   & $s_{T,C}$ & $d_{T,C}/h$       \\
\hline
\end{tabular} 
\label{SensApp}
\end{center}
\end{table}

Interestingly, for a fixed wavelength $\lambda$,  the phase $\phi(s_{C}, d_C, s_{T}, d_T, x_C, x_T )$ is determined by the pinhole separations $d_C$ and $d_T$ weighted either  by the average transverse positions $s_j$ of the two pinholes divided by $h$, or by
the detection angles $x_j/f $ evaluated with respect to the optical axis. 
Therefore, the correlation measurement of Eq. (\ref{Deltan4}) is sensitive to the position and the transverse structure of the two masks.

\begin{figure}[h!] 
\centering\includegraphics[width=11cm]{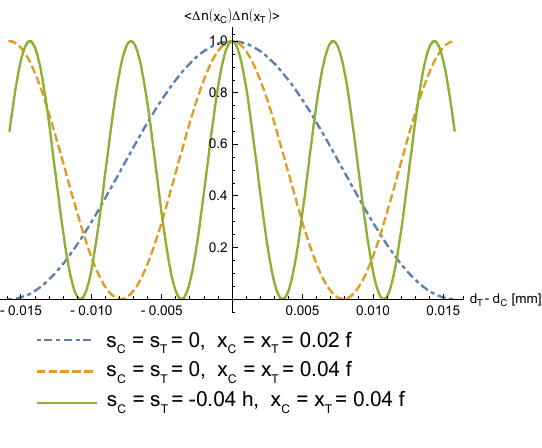}
\caption[]{
 {Simulation of the measurement} of  the stretching/shrinking $d_C-d_T$ of one mask with respect to the other in the setup of Fig. \ref{C-NotPoP1a} with $z=500 mm$ and $f=100mm$. The source 
 {is assumed to have a constant profile, with size $a=2mm$, and wavelength $\lambda=632 nm$, so that the coherence length is $l_{coh} = \lambda z/a =0.158mm$. When the two pinholes in each mask are placed symmetrically with respect to the optical axis ($s_C=s_T=0$), the observable effect of small variations in $d_C-d_T$   is enhanced when the transverse position $x_C = x_T$ of the two detectors is increased, as demonstrated by the dashed (yellow) curve as compared to the dash-dot (blue) one. A further enhancement is obtained  by displacing equally both masks with respect to the optical axis in the opposite direction of the detectors, as demonstrated by the continuous (green) curve.}} \label{ResMag}
\end{figure}

In Table \ref{SensApp}, we consider five different experimental scenarios exploiting  correlation measurement for monitoring small variations in: 
 {(i)} the difference $d_C - d_T$ of the two pinhole separations, 
 {(ii)} the sum $d_C + d_T$ of the two pinhole separations,
 {(iii)} the pinhole separation $d_j$ in one mask $j =C,T$ if the separation in the other mask is fixed, 
 {(iv)} the relative position $s_T - s_C$ of the masks, 
 {(v)} the transverse position $s_j$ of one mask $j=C,T$ if the position of the other mask is fixed. 
Interestingly, as reported in the third column of Table \ref{SensApp}, in all scenarios it is possible to 
  {increase the precision} of the measurement  without {either} increasing the frequency of the light or using entanglement{: the trick is to employ the remaining spatial parameters to ``magnify'' the effect of the variation of the spatial parameter to be monitored. An analysis of the sensitivity of this technique in terms of the number of resources is beyond the scope of this paper and will be addressed in future research \cite{GioLloMac}.}

In Fig. \ref{ResMag}, we depict the first experimental scenario reported in Table \ref{SensApp}, where, for equal transverse positions $s_C = s_T$ of the two masks, the correlation measurement at equal detector positions $x_C = x_T$ is sensitive to the stretching/shrinking  $d_c-d_T$ of one mask with respect to the other. 
 {In the simple} case where the two pinholes in each mask are placed symmetrically with respect to the optical axis ($s_C =s_T=0$), 
 {the effect  of small variations in $d_C-d_T$ can be magnified} by moving both detectors at larger angles $x_C/f$ with respect to the optical axis (dashed yellow curve). A further enhancement can be obtained by displacing both masks equally with respect to the optical axis, but in the opposite direction of the detectors (green continuous curve).

\begin{table}
\begin{center}
\caption{Summary of the 
 experimental conditions for 
 {characterizing two remote double-pinhole masks by measuring in the setup in Fig. \ref{C-NotPoP1a} the period} of the second order interference  pattern given by  Eq. (\ref{Deltan4}).}
\begin{tabular}{|l|l|l|}
\hline
    Experimental conditions    & Experimental    &   Period of the  interference \\   in addition to  
 {Eqs.} (\ref{Limit2b}) and (\ref{Limit2a})& variable&  pattern $\langle \Delta n (x_C) \Delta n (x_T) \rangle$ \\ 
\hline
{(i)} $x_T=x_C$ & $x_C$ & $\lambda f / ({d_C - d_T})$          \\
{(ii)} $x_T=-x_C$   & $x_C$  &$ \lambda  f / ({d_T + d_C})$     \\
{(iii)} $s_{T}=s_{C}$, $x_C = x_T=0$  & $d_T-d_C$  &  $\lambda  h / {s_{C} }$  \\
{(iv)} $s_{T}=-s_{C}$, $x_C = x_T=0$ & $d_T+d_C$  & $\lambda  h/ {s_{C} }$     \\
{(v)} $d_T=d_C$, $x_C = x_T=0$ & $s_{T}-s_{C}$  & $\lambda  h/ {d_C}$     \\
\hline
\end{tabular}  
\label{Tab2}
\end{center}
\end{table}

Based on Eqs. (\ref{Deltan4}) and (\ref{phiP}){,} the transverse structure of the two masks can also be retrieved, indirectly, by measuring the period of the second-order  interference pattern $\langle \Delta n (x_C) \Delta n (x_T) \rangle$ obtained in  the experimental scenarios reported in Table \ref{Tab2}.
For example, by performing correlation measurements at both equal and opposite positions with respect to the optical axis (first and second experimental scenarios, respectively, in Table \ref{Tab2}) it is possible to retrieve the pinhole separations $d_C$ and $d_T$ in each mask.

Interestingly, the 
  sensing capabilities of the present interferometric technique 
 have currently no counterparts in the temporal domain \cite{Tam-Sei}.

\section{Simulation of quantum logic gates}\label{SimCNOT}

In this section, we show that quantum logic operations can be simulated  by using the spatial interference effect described so far. In particular, we address the simulation of  a controlled-$U_{\phi}$ gate, with $U_{\phi}$ described by the matrix \cite{nielsen}
\begin{eqnarray}\label{Ucontrol}
U_{\phi}\defeq
\begin{pmatrix}
0  & e^{i \phi}\\
e^{i \phi}  & 0
\end{pmatrix}.
\end{eqnarray}

Let us start by describing  a \textit{genuine} controlled-$U_{\phi}$ gate. Given two-qubit input states $\ket{\phi_{\Ccurly}}_{\Ccurly}\ket{\phi_{\Tcurly}}_{\Tcurly}$, where 
\begin{equation}
\ket{\phi_{\Ccurly}}_{\Ccurly}\defeq\cos\phi_{\Ccurly}\ket{H}_{\Ccurly}+\sin\phi_{\Ccurly}\ket{V}_{\Ccurly},
\end{equation}
and
\begin{equation}
\ket{\phi_{\Tcurly}}_{\Tcurly}\defeq\cos\phi_{\Tcurly}\ket{H}_{\Tcurly}+\sin\phi_{\Tcurly}\ket{V}_{\Tcurly},
\end{equation}  
the controlled-$U_{\phi}$ gate operates on the input states, by giving the following output entangled state \cite{nielsen}
\begin{eqnarray}
\ket{\psi}_{}&=&\cos\phi_{\Ccurly}\ket{H}_{C}\ket{\phi_{\Tcurly}}_{T}+e^{i \phi}\sin\phi_{\Ccurly}\ket{V}_{C}\ket{\phi_{\Tcurly}^{(F)}}_{T},
\end{eqnarray}
where 
\begin{equation}
\ket{\phi_{\Tcurly}^{(F)}}_{T}\defeq\sin\phi_{\Tcurly}\ket{H}_{T} + \cos\phi_{\Tcurly}\ket{V}_{T}.
\end{equation}
The  polarization-dependent  joint detection  probability associated with the state $\ket{\psi}_{}$ is \cite{nielsen}
\begin{eqnarray}\label{PCU}
P_{U_{\phi}}\defeq\abssq{\braket{\theta_{C},\theta_{T}|\psi}_{}}
=\Big|\cos\phi_{\Ccurly}\cos\theta_{C} \cos\left(\phi_{\Tcurly}-\theta_{T}\right)
+e^{i \phi}  \sin\phi_{\Ccurly}\sin\theta_{C} \sin\left(\phi_{\Tcurly}+\theta_{T}\right) \Big|^{2}.
\end{eqnarray} 
In particular, for $\phi =0$, the controlled-$U_{\phi}$ gate reduces to a CNOT gate \cite{nielsen} and the  polarization-dependent joint detection probability in Eq. (\ref{PCU}) becomes
\begin{eqnarray}\label{PCNOT}
P_{\text{CNOT}}\defeq\abssq{\braket{\theta_{C},\theta_{T}|\psi}_{}}
=\left|\cos\phi_{\Ccurly}\cos\theta_{C} \cos\Big(\phi_{\Tcurly}-\theta_{T}\right)+ \sin\phi_{\Ccurly}\sin\theta_{C} \sin\left(\phi_{\Tcurly}+\theta_{T}\right) \Big|^{2}. 
\end{eqnarray}

\begin{figure} [h!]
\centering\includegraphics[width=14cm]{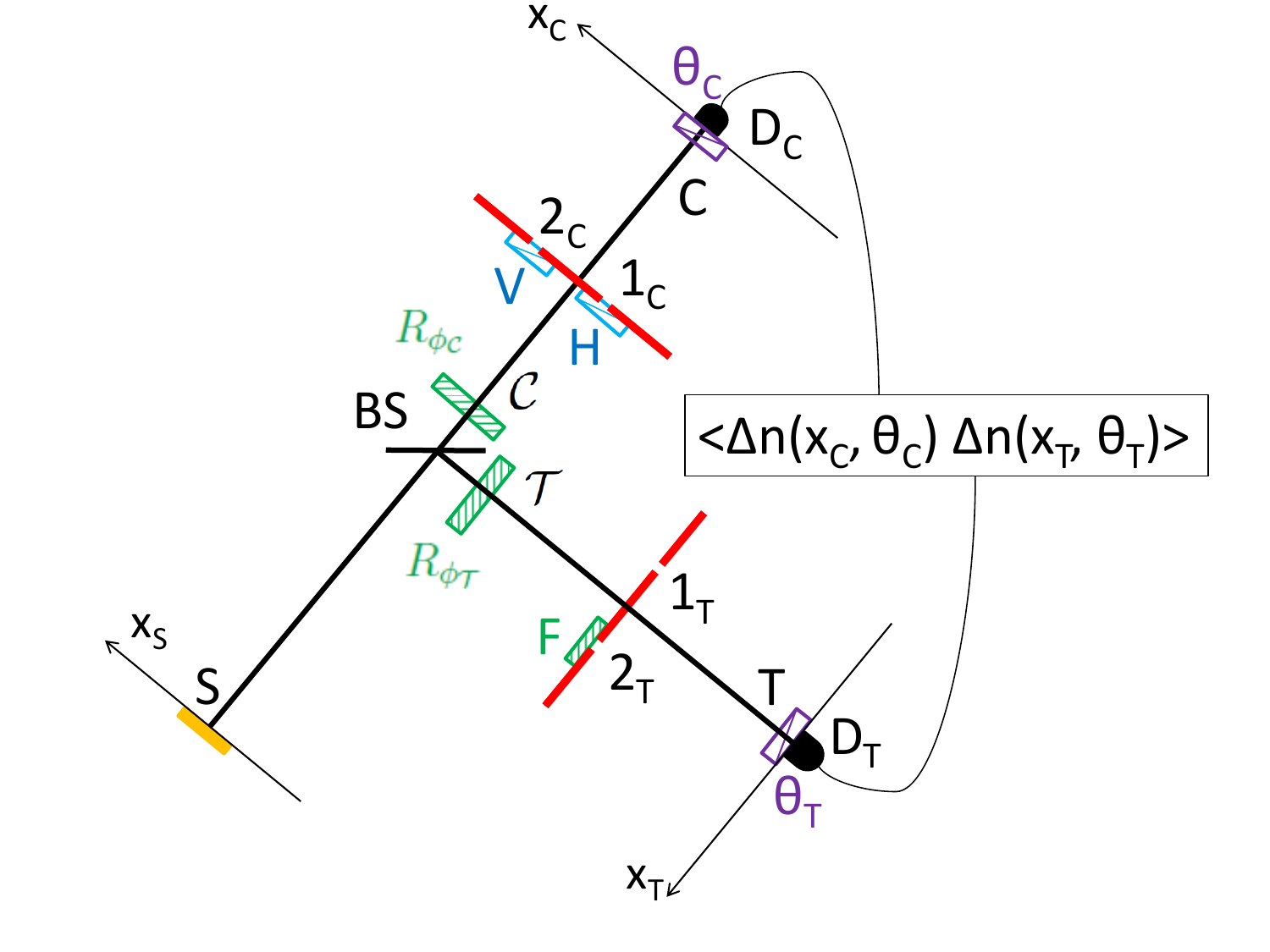}
\caption[]{Interferometer for the simulation of controlled-$U_{\phi}$ gates, with $U_{\phi}$ defined in Eq. (\ref{Ucontrol}). In the first part of the interferometer, the initial polarization state of the light is prepared. The second part, from the ports $\Ccurly$ and   $\Tcurly$ to the ports $C$ and $T$, respectively, performs a polarization-dependent transformation. Correlation measurements in the fluctuations of the number of photons are performed at the interferometer output. $R_{\phi_\Ccurly}$ and $R_{\phi_\Tcurly}$  are two half-wave plates that rotate the polarization of the angles $\phi_C$ and $\phi_T$, respectively; $F$ is a half-wave plate  implementing a  flip from the horizontal (H) polarization to the vertical (V) polarization and vice versa; $H$, $V$, $\theta_C$ and $\theta_T$  represent the polarization directions of the corresponding polarizers.} \label{C-NotPoP1b}
\end{figure}

In order to simulate a controlled-$U_{\phi}$ gate we propose in Fig. \ref{C-NotPoP1b} a modification of the interferometer in Fig. \ref{C-NotPoP1a}. The interferometer consists of  three parts: the first one prepares the initial polarization state in the ``control'' input  port $\Ccurly$ and in the ``target'' input port $\Tcurly$;  the second one implements polarization transformations along the control and target output channels; the final part consists of the measurement process.

In the first part of the setup, the H-polarized chaotic light impinges on a balanced non-polarizing beam splitter and then propagates through two half-wave plates $R_{\boldsymbol\phi_{\Ccurly}}$ and  $R_{\boldsymbol\phi_{\Tcurly}}$. 
 
The second part of the setup consists of a ``control'' path, connecting the ports $\Ccurly$ and $C$,  and a  ``target'' path, connecting the ports $\Tcurly$ and $T$. Similar to the setup in Fig. \ref{C-NotPoP1a}, both in the control and in the target paths  light goes through identical two-pinhole masks. However, in the control path, two polarizers oriented along the $H$ and $V$ directions are placed just before pinholes $1_C$ and $2_C$, respectively, while in the target path a half-wave plate  oriented at $\pi/4$  is placed just before the pinhole $2_T$.

Let us now describe the detection process. A polarizer, oriented along the direction $\boldsymbol{\theta}_{d}\defeq\left( \cos\theta_{d}\quad\sin\theta_{d}\right)^T$, with $d=C,T$, is placed in front of each detector. A polarization-dependent correlation measurement between the fluctuations of the number of photons $\Delta n(x_C, \theta_{C})$ and $\Delta n(x_T, \theta_{T})$, detected, respectively, by $D_C$ and $D_T$, is then performed.

As shown in Appendix \ref{app2}, if the conditions {in Eqs.} (\ref{Limit2b}) and (\ref{Limit2a}) are satisfied, in the paraxial approximation the correlation between the fluctuations of the number of photons is proportional to the joint detection probability typical of a controlled-$U_{\phi}$ gate, namely  
\begin{eqnarray}\label{G1effFinP} 
\avg{\Delta n_{}({x_C, \theta}_{C}) \Delta n_{}(x_T, {\theta}_{T})}_{}\propto\Big| G_{{1_C,1_T}}^{(1)}({x_C, \theta}_{C}, x_T, {\theta}_{T})+G_{{2_C, 2_T}}^{(1)}(x_C, {\theta}_{C}, x_T, {\theta}_{T})\Big|^2 \propto P_{U_{\phi}},
\end{eqnarray}
with $\phi$  defined in Eq. (\ref{phiP}). However, differently from the setup in Fig. \ref{C-NotPoP1a}, the two interfering contributions $G^{(1)}_{1_C,1_T}$ and $G^{(1)}_{2_C,2_T}$, associated with the propagation through the two pairs of pinholes ($1_C, 1_T$) and ($2_C, 2_T$), are polarization dependent. In particular: 
\begin{enumerate}
\item the control path $1_C$, associated with the polarization mode H, is correlated with the target path $1_T$, where the polarization is not modified;
\item the control path $2_C$, associated with the polarization mode V, is correlated with the target path $2_T$, where the polarization is flipped from H to V, and vice versa. 
\end{enumerate}
Interestingly, the resulting second-order interference pattern is proportional to the probability $P_{U_{\phi}}$  associated with a controlled-$U_\phi$ gate, with $\phi$ defined in Eq. (\ref{phiP}). In particular, when
\begin{eqnarray} \label{Phi<<} 
&&|\phi(s_{C}, d_C, s_{T}, d_T, x_C, x_T )|\ll1,
\end{eqnarray}
Eq. (\ref{G1effFinP}) reduces to
\begin{eqnarray}
&&\avg{\Delta n_{}({x_C, \theta}_{C}) \Delta n_{}(x_T, {\theta}_{T})}_{}\propto P_{CNOT},
\end{eqnarray}
with $P_{CNOT}$  defined in Eq. (\ref{PCNOT}), leading to the \textit{simulation} of a CNOT gate operation without recurring to any entanglement processes. 

Based on Eq. (\ref{phiP}), the condition reported in  Eq. (\ref{Phi<<}) can be experimentally obtained, for example, by performing the detections at equal 
 {positions} $x_C=x_T$ with the pinholes in the two masks placed in the same position with respect to the optical axis ($d_C = d_T$, $s_C= s_T$).

By using a generalized N-port beam splitter and N double-slit masks, the scheme in Fig. \ref{C-NotPoP1b} can be generalized for the simulation of interference features typical of N-order entangled correlations.

\section{Discussions}

Based on the setup in Fig. \ref{C-NotPoP1a}, we have theoretically demonstrated a second-order spatial interference effect  between two pairs of disjoint but correlated paths. The two interfering paths are associated with the pairs of remote pinholes $1_C, 1_T$ and $2_C, 2_T$. Interestingly, such interference exists even if the  pinholes in each mask are separated by a distance much larger than the transverse coherence length $l_{coh}$ of the source. 
In fact, the interference between the pairs of paths ($1_C, 1_T$) and ($2_C, 2_T$) arises from the correlation between the two disjoint paths going through pinholes  $1_C, 1_T$ and $2_C, 2_T$, respectively; in fact, the transverse distance between the  two pinholes $1_C$ and $1_T$ (or $2_C$ and $2_T$) is smaller than the transverse coherence length of the source. This is not the case for the other two possible pairs of paths, ($1_C, 2_T$) and ($2_C, 1_T$), which therefore cannot contribute to the interference. This phenomenon,substantially  different from all second-order interference phenomena based on multiple chaotic sources  \cite{Agaf, Pearce2015, SVS, Oppel2012}, thus provides a deeper  understanding of the physics of multi-path interference and spatial coherence.

Furthermore, we have demonstrated that this spatial interference effect  has interesting potential applications for sensing of remote objects in the absence of first-order coherence. In particular, we have shown that information about both the  transverse structure and  the relative position of two remote double-pinhole masks is encoded within the relative phase between the two interfering pairs of optical paths (Eq. (\ref{phiP})). These spatial parameters can be retrieved  through the measurement of the period of the second order interference pattern given by Eq. (\ref{Deltan4}) (Table \ref{Tab2}). Remarkably, 
{the effect produced on the correlation measurement by small variations of these spatial parameters can be enhanced without increasing the frequency of the light, as demonstrated in Table \ref{SensApp} and in the example in Fig. \ref{ResMag}}. This may lead to novel applications in sensing  biological samples  without exposure to high-frequency light \cite{crespi2012}. Moreover, this technique can be applied independently of the distances between the two masks and the source and between the masks and  the corresponding detectors. Therefore, this effect can be potentially employed for monitoring the relative spatial structure and position of distant objects.

In addition, we have demonstrated how to exploit this novel spatial interference phenomenon for simulating entanglement correlations, including the simulation of a CNOT gate (Fig. \ref{C-NotPoP1b}). This technique can be used, 
 to simulate typical interference features of high-order entanglement correlations with potential applications in novel optical algorithms \cite{tamma_analogue_2015-1, tamma_analogue_2015, PRARapidFact, JOMFact, TamAll, Wolk2011}.

In conclusion the proposed spatial interference effect  provides a deeper understanding of the physics of  spatial coherence and multi-photon interference, and can naturally lead to novel interferometric techniques for sensing distant objects and simulating small-scale quantum circuits. 
  {This interference phenomenon may also be extended to atomic interferometers with thermal bosons, for example, to measure the effect of external forces (e.g., gravity) on bosons of given mass in remote spatial regions.}

\appendix

\section{Green's propagator for the setup in Fig. \ref{C-NotPoP1a}}\label{appGF}

Given the optical setup in Fig. \ref{C-NotPoP1a} we calculate here the Green's propagator $g\{ \kappa; S, x_d\}$, associated with the $x$ component $\kappa$ of transverse wave-vector, from the source $S$ with amplitude profile $A(x_S)$ to the detector transverse position $x_d$, with $d= C,T$. In particular, we obtain \cite{D'AnShi, ShihBook, Rubin}
\begin{eqnarray}\label{GreFunn0}
g\{ \kappa; S, x_d\}&=&\frac{1}{\sqrt{2}}e^{i \varphi(d)}\int d x_S d x_M  A(x_S) M(x_M) e^{i \kappa x_S}  \left\{\frac{- i \omega}{2 \pi c} \frac{e^{ i \omega z/c}}{z} \Gmc(|x_S-x_M|)_{[\omega/(c z)]}\right\}\nonumber\\ &\times& \left\{ \frac{- i \omega}{2 \pi c} \frac{e^{ i \omega f/c}}{f}  \Gmc(|x_M-x_d|)_{[\omega/(c f)]}\right\},
\end{eqnarray}
where $\omega$ is the frequency of the light,
\begin{eqnarray}\label{MaskFunction}
M(x_M)\defeq \sum_{x_p} \delta(x_M - x_p)
\end{eqnarray}
is the mask transfer function, defined by the transverse position $x_p$ of the pinholes $p=1_C, 2_C$ for the upper mask and $p=1_T, 2_T$ for the lower mask,
\begin{eqnarray}\label{FrPrPhFa}
\Gmc(|\alpha|)_{[\beta]} \defeq e^{i \frac{\beta}{2} |\alpha|^2}
\end{eqnarray}
 is the Fresnel propagator, and the factor $\frac{1}{\sqrt{2}}e^{i \varphi(d)}$ takes into account the propagation through the beam splitter, with $\varphi(C)=0$ for the transmitted beam and $\varphi(D)=\pi/2$  for the reflected beam.

By using  the definition (\ref{FrPrPhFa}) and the property 
\begin{eqnarray}
\Gmc(|\alpha+\alpha'|)_{[\beta]}&=&\Gmc(|\alpha|)_{[\beta]}\Gmc(|\alpha'|)_{[\beta']} e^{i \beta \alpha \alpha'} 
\end{eqnarray}
of the Fresnel propagator, and the mask transfer function  in Eq. (\ref{MaskFunction}), Eq. (\ref{GreFunn0}) becomes
\begin{eqnarray}\label{GreFunn2}
g\{ \kappa; S, x_d\}= \sum_{p=1_d,2_d} B_j(x_d)   \int d x_S  A(x_S)   \Gmc(|x_S|)_{[\omega/(c z)]} e^{ i[\kappa -\omega  x_p  /(zc)] x_S}, 
\end{eqnarray}         
where
\begin{eqnarray}\label{GreFunn21}
 B_p(x_d) \defeq -\frac{1}{\sqrt{2}}\left(\frac{\omega}{2 \pi c}\right)^2 \frac{e^{ i[\varphi(d)+ \omega (z+f)/c]}}{z f}\Gmc(|x_d|)_{[\omega/(c f)]} \Gmc(|x_p|)_{[\omega/(c h)]}  e^{ -i \omega  x_d x_p /(fc)} .
\end{eqnarray}         

The Green's function in Eq. (\ref{GreFunn2}) can be finally written as the sum
\begin{eqnarray}\label{GreFunn22a}
g\{ \kappa; S, x_d\}&=&  \sum_{p=1_d,2_d} g_p\{ \kappa; S, x_d\}, 
\end{eqnarray}         
of the two Green's propagators
\begin{eqnarray}\label{GreFunn2b}
g_p\{ \kappa; S, x_d\}\defeq  B_p(x_d) \int d x_S  A(x_S)   \Gmc(|x_S|)_{[\omega/(c z)]}  e^{ i[\kappa -\omega  x_p  /(zc)] x_S},
\end{eqnarray}
from the source S to the detector position $x_d$, with $d=C,T$, through the pinhole located in $x_p$,  with $p=1_d, 2_d$.

\section{Correlation measurement for the setup in Fig. \ref{C-NotPoP1a}}\label{app1}

In the present appendix we present a detailed derivation of the correlation in the fluctuation of the numbers of photons in Eqs. (\ref{Deltan3}) and (\ref{Deltan4}) measured at the output of the setup in Fig.\ref{C-NotPoP1a}.

By substituting in Eq. (\ref{G1n}), the definition of the electric field operator (Eq. (\ref{Elecn})  with the Green's propagator in Eq. (\ref{GreFunn22a}), we obtain the first order correlation function
\begin{eqnarray}
G^{(1)} (x_{C}, x_T)=\sum_{\substack{p=1_C, 2_C \\ q=1_T, 2_T}} |K|^2 Tr\Big{[}\density{H}\int  d \kappa  d \kappa'   g_p^*\{ \kappa; S, x_C\}  g_q\{ \kappa'; S, x_T\} \hat{a}^{{\dagger}}_S(\kappa) \hat{a}_S(\kappa') \Big{]}.
\end{eqnarray}
This expression corresponds to the sum
\begin{eqnarray}\label{Deltan10md} 
G^{(1)}_{} (x_{C}, x_T)= \sum_{\substack{p=1_C, 2_C \\ q=1_T, 2_T}}G^{(1)}_{p, q} (x_{C}, x_T). 
\end{eqnarray}
of the four contributions
\begin{eqnarray}\label{Deltan10} 
G^{(1)}_{p, q} (x_{C}, x_T) \defeq |K|^2 Tr \Big{[} \density{H}\int  d \kappa  d \kappa'  g_p^*\{ \kappa; S, x_C\}  g_q\{ \kappa'; S, x_T\} \hat{a}^{{\dagger}}_S(\kappa) \hat{a}_S(\kappa') \Big{]},
\end{eqnarray}
from the corresponding four pairs of optical paths $(p,q) = (1_C,1_T), (2_C,2_T), (1_C,2_T), (2_C, 1_T)$.

By using the  property of chaotic sources  \cite{Mandel1995}
\begin{eqnarray}\label{TrThSo}
Tr\left[\density{} a^{{\dagger}}(\kappa) a(\kappa')\right]=  \braket{n_{\kappa}} \delta({\kappa-\kappa'}),
\end{eqnarray}
where  the average photon number $\braket{n_{\kappa}}$ in the mode $\kappa$ is assumed to be constant, and  the Green's propagators in  Eq. (\ref{GreFunn2b}), Eq. (\ref{Deltan10}) reduces to  
\begin{eqnarray}\label{G1eff10} 
{ G^{(1)}_{p, q} (x_{C}, x_T) = K'  B_p^*(x_C)  B_q(x_T) FT \left\{ | A(x_S)|^2 \right\} \left[\omega (x_p-x_q) /(2 \pi c z) \right],}
\end{eqnarray}
where $K' \defeq |K|^2 \braket{n_{\kappa}}$ and  $FT \left\{ | A(x_S)|^2 \right\} \left[\chi \right]$ represents the Fourier transform of the source intensity profile, calculated in $\omega (x_p-x_q) /(2 \pi c z)$.
If $|x_p-x_q| \gg  l_{coh}$, this Fourier transform is approximately zero, so that no contribution to the correlation function in Eq. (\ref{Deltan10md}) arises from the pair of paths $(p,q)$. On the contrary the pair of path $(p,q)$ gives its maximum contribution if  $|x_p-x_q|\ll l_{coh}$.
This implies that, in the conditions {given in Eqs.} (\ref{Limit2b}) and (\ref{Limit2a}), the correlation function in Eq. (\ref{Deltan10md}) reduces to the sum 
\begin{eqnarray}
G^{(1)} (x_{C}, x_T)&=&G_{1_C,1_T}^{(1)} (x_{C}, x_T)+G_{2_C,2_T}^{(1)} (x_{C}, x_T)
\end{eqnarray}
of the only two contributions associated with the pairs of paths $(1_C,1_T)$ and $(2_C,2_T)$. By substituting this expression in Eq. (\ref{Deltan}), we obtain the correlation in the photon-number fluctuations $\avg{\Delta n(x_C) \Delta n(x_T)}_{}$ in Eq. (\ref{Deltan3}).

By using the conditions {in Eqs.} (\ref{Limit2b}) and (\ref{Limit2a}) and Eq. (\ref{G1eff10}), Eq. (\ref{Deltan3}) can be written explicitly as
\begin{eqnarray} 
\avg{\Delta n(x_C) \Delta n(x_T)}_{}&=&\big{|} K'  FT \Big\{ |A(x_S)|^2\Big\}(0)  \big{[} B_{1_C}^*(x_C)  B_{1_T}(x_T) + B_{2_C}^*(x_C)  B_{2_T}(x_T) \big{]}\big{|}^2.
\end{eqnarray}
By inserting the expressions in Eq. (\ref{GreFunn21}) with the definition of the Fresnel propagator (Eq. (\ref{FrPrPhFa})), 
we finally obtain
\begin{eqnarray}
\avg{\Delta n(x_C) \Delta n(x_T)}_{}&=&K''  \big{|}  e^{-i \omega/(2 c h) (x_{1_C}^2 - x_{1_T}^2)} e^{i \omega/(c f)(x_C x_{1_C}-x_T x_{1_T})}\nonumber\\&+& e^{-i \omega/(2 c h) (x_{2_C}^2 - x_{2_T}^2)}  e^{i \omega/(c f)(x_C x_{2_C}-x_T x_{2_T})} \big{|}^2, 
\end{eqnarray}
with $K'' \defeq \big{|}\left(i/2\right)\left[1/\left(z f\right)\right]^2 K'  \left[\omega/\left(2 \pi c\right)\right]^4 FT \left\{ | A(x_S)|^2\right\}(0)\big{|}^2$, which reduces easily to Eq. (\ref{Deltan4}).

\section{Correlation measurement for the setup in Fig. \ref{C-NotPoP1b}}\label{app2}

In the present appendix we derive the correlation in the fluctuations of the number of photons in  Eq. (\ref{G1effFinP}), measured at the output of the interferometer in Fig. \ref{C-NotPoP1b} for arbitrary polarization angle $\theta_C$ and $\theta_T$.
For a H-polarized quasi-monochromatic 1-dim chaotic source (thermal state 
  {$\density{H}$} in Eq. (\ref{eq:Density_op_Thermal_B_Two_Mode0})), this correlation is given by \cite{Glauber1963}
\begin{equation}
\avg{\Delta n_{}({x_C, \theta}_{C}) \Delta n_{}({{x_T, \theta}_{T}})}_{}=\abssq{G^{(1)}_{}({x_C, \theta}_{C}; {x_T}, {\theta}_{T})}
\end{equation}
where

{\begin{eqnarray}\label{G1eff0S} 
 G_{}^{(1)}( x_C, \theta_C; x_T, \theta_T) &=& Tr\left[\density{H} \ \hat{\mathcal{E}}_{C, S}^{{(-)}}(x_C) \ \hat{\mathcal{E}}_{T, S}^{{(+)}}(x_T)\right]\nonumber
\\ &=&  {K' \int  d \kappa    L^*_{C}(\kappa)   L_{T}(\kappa)},
\end{eqnarray}
{is the first-order correlation function determined by the field operator}
\begin{eqnarray} 
\hat{\mathcal{E}}_{d, S}^{{(+)}}(x_d)&{\defeq}& K \int  d \kappa \ e^{- i \omega t} L_{d}(\kappa) \ \hat{a}_S^{(H)}(  \kappa), 
\end{eqnarray}
{with $d= C,T$, where K is a constant factor, $K' \defeq \big{|} K \big{|} ^2 \braket{n_{\kappa}}$ and 
\begin{eqnarray} 
 L_C (\kappa) &\defeq& \frac{1 }{\sqrt{2}} \big{[}  g_{1_C}\{\kappa; S, x_C\} \cos{\theta_C}\cos\phi_{\Ccurly}  +  g_{2_C}\{\kappa; S, x_C\} \sin{\theta_C} \sin\phi_{\Ccurly} \big{]} \label{ACS}, \\
 L_T (\kappa) &\defeq& \frac{i }{\sqrt{2}} \big{[}   g_{1_T}\{\kappa; S, x_T\} \cos(\theta_{T}-\phi_{\Tcurly})  + g_{2_T}\{\kappa; S, x_T\} \sin(\theta_{T}+ \phi_{\Tcurly}) \big{]}  \label{ATS}
\end{eqnarray}
are the effective propagation functions.}
By substituting the expressions  in Eqs. (\ref{ACS}) and (\ref{ATS}), the correlation function in Eq. (\ref{G1eff0S})} becomes
\begin{eqnarray}\label{G1eff1S} 
 G_{}^{(1)}( x_C, \theta_C; x_T, \theta_T)= \frac{i }{2} K'  \int  d \kappa \Big{[}\cos{\theta_C}\cos\phi_{\Ccurly}\cos(\theta_{T}-\phi_{\Tcurly})   g^*_{1_C}\{\kappa; S, x_C\}   g_{1_T}\{\kappa; S, x_T\} \nonumber\\
+ \sin{\theta_C} \sin\phi_{\Ccurly} \sin(\theta_{T} + \phi_{\Tcurly})  g^*_{2_C}\{\kappa; S, x_C\}  g_{2_T}\{\kappa; S, x_T\} \nonumber\\
 -  \cos{\theta_C}\cos\phi_{\Ccurly}  \sin(\theta_{T} + \phi_{\Tcurly})  g^*_{1_C}\{\kappa; S, x_C\}  g_{2_T}\{\kappa; S, x_T\} \nonumber\\
 -\sin{\theta_C} \sin\phi_{\Ccurly} \cos(\theta_{T} - \phi_{\Tcurly}) g^*_{2_C}\{\kappa; S, x_C\}  g_{1_T}\{\kappa; S, x_T\}\Big{]}.
\end{eqnarray}
By using the result in Eq. (\ref{G1eff10}) and by applying the conditions {given in Eqs.} (\ref{Limit2b}) and (\ref{Limit2a}) in an analogous way as in Appendix \ref{app1}, Eq. (\ref{G1eff1S}) reduces to Eq. (\ref{G1effFinP}).

\section*{Funding}
M.C. acknowledges a Ph.D. studentship from the University of Bari.   M.C., M. D. and A.G. acknowledge funding from P.O.N. RICERCA E COMPETITIVITA' 2007-2013 - Avviso n. 713/Ric. del 29/10/2010, Titolo II - ``Sviluppo/Potenziamento di DAT e di LPP'' (project n. PON02-00576-3333585). V.T. acknowledges the support of the German Space Agency DLR with funds provided by the Federal Ministry of Economics and Technology (BMWi) under grant no. DLR 50 WM 1556

\section*{Acknowledgments}
M.C. is thankful for the hospitality of Prof. W. P. Schleich during his visit in the summer of 2015 at the Institute of Quantum Physics, Ulm University. M.C. and V.T. would also like to thank J. Seiler for useful discussions during this period.

\end{document}